\documentclass[preprintnumbers,a4paper,twocolumn,10pt,showpacs,prl]{revtex4}%
\usepackage{graphicx}
\usepackage{dcolumn}
\usepackage{bm}
\usepackage{latexsym,amsthm,amssymb}
\usepackage{booktabs,threeparttable}
\usepackage{makeidx}
\usepackage{amsmath}
\usepackage{amsfonts}
\usepackage{amssymb}%
\providecommand{\U}[1]{\protect\rule{.1in}{.1in}}
\newcommand{\bean}{\begin{eqnarray*}}
\newcommand{\eean}{\end{eqnarray*}}
\newcommand{\be}{\begin{equation}}
\newcommand{\ee}{\end{equation}}
\newcommand{\ba}{\begin{array}}
\newcommand{\ea}{\end{array}}
\newcommand{\bea}{\begin{eqnarray}}
\newcommand{\eea}{\end{eqnarray}}

\begin{document}
\title{Quantum critical surface of the zigzag spin chain under magnetic field:
Application to superconducting quantum dots}
\author{Meihua Chen$^{1}$, Sujit Sarkar$^{2}$, C. D. Hu$^{1}$ }
\affiliation{$^{1}$Physics Department, National Taiwan University, Taipei, R. O. C }
\affiliation{$^{2}$Poornaprajna Institute of Scientific Research, 4 Sadashivanagar,
Bangalore- 5600 80, India}
\date{\today }

\begin{abstract}
\noindent We analyze the exact ground state of XXZ zigzag spin chain with
applied magnetic field and find the quantum critical surface. Using the
theorem of positive semi-definite matrix, we can prove that the ground states
for a specific region, are fully polarized state and one magnon states. With
Bethe ansatz, we argue that this is the quantum critical surface in all cases.
A first in the literature, we derive the analytical expression of quantum
critical surface for superconducting quantum dots array in the presence of
gate voltage.

\end{abstract}
\maketitle

\noindent Low dimensional spin systems has been an active field of research
\cite{QM04} for many year. Bethe ansatz has been used widely to solve many new
families of integrable Hamiltonians of one dimensional systems
\cite{Bethe31,Mul98,Tak99,Zvy01}. A quantum inverse scattering method based on
Bethe ansatz has also been developed for dealing with the two-dimensional
models \cite{Inverse93}. Another method which also enables one to find the
exact solutions of various models is the matrix product method
\cite{Zitt91,VBS95,KK94, Gu96} . For example, it is related to the
Affleck-Kennedy-Lieb-Tasaki(AKLT) state \cite{KK94,AKLT87}. In this work we
are going to apply these two methods to the zigzag spin chain model.\newline A
zigzag spin chain is an one dimensional system of spins with exchange
interactions with their nearest (NN) and next-nearest neighbors (NNN), denoted
by $J_{1}$ and $J_{2}$ respectively. It is a frustrated system if $J_{1}%
J_{2}<0$ or both of them are positive. In a zigzag spin chain the situation is
not clear. Majumdar and Ghosh \cite{MG69} found that the degenerate dimer
states\ in which neighboring sites form singlets (MG state) are the exact
ground states at $\frac{J_{1}}{J_{2}}=2$. At $\frac{J_{1}}{J_{2}}=-4$, Hanada
et.al. \cite{Nat88} found that the uniformly distributed resonate valence bond
(UDRVB) state and the fully polarized (FP) state are degenerate ground states.
There are also a few other important theoretical works in zigzag spin chain
\cite{W08,A95,RS97,SS02}. The zigzag spin chain can be realized in physical
systems, such as the oxides, Li$_{2}$CuO$_{2}$ (multiferroic) \cite{Ma98} and
SrCuO$_{2}$ \cite{Uch97}.\newline In this letter, we find a quantum critical
surface on which the FP state is degenerate with the one-magnon states, along
with an application to superconducting quantum dots system. Our calculation is
based on the concept that the local ground states in a matrix product form can
be extended to global ground states. \newline\textbf{Model Hamiltonian:} We
begin with the Hamiltonian of a zigzag spin chain with applied magnetic
field,
\begin{align*}
H  &  =\sum_{i}J_{1}[\Delta_{1}s_{i}^{z}s_{i+1}^{z}+\frac{1}{2}(s_{i}%
^{+}s_{i+1}^{-}+s_{i}^{-}s_{i+1}^{+})]\\
&  +\sum_{i}J_{2}[\Delta_{2}s_{i}^{z}s_{i+2}^{z}+\frac{1}{2}(s_{i}^{+}%
s_{i+2}^{-}+s_{i}^{-}s_{i+2}^{+})]+B\sum_{i}s_{i}^{z}%
\end{align*}
where $J_{1}$ and $J_{2}$ are the NN and, NNN interaction, $\Delta_{1}%
(\Delta_{2})$ is the anisotropy in $z$ axes for NN(NNN) interaction and $i$ is
the label of the lattice site.
Furthermore, we use the periodic boundary condition and hence we have the
coupling between $\vec{s}_{N-1},\vec{s}_{N}$ and $\vec{s}_{1},\vec{s}_{2}$. We
can dissect the Hamiltonian into many local Hamiltonians. Each local
Hamiltonian contains three spins. Thus, the original Hamiltonian can be
written as
\begin{equation}
H=\sum_{i}h_{i,i+1,i+2}=\sum_{i}(h_{i,i+1,i+2}^{0}+h_{i,i+1,i+2}^{mag}).
\end{equation}
where
$h_{i,i+1,i+2}^{0}=\frac{f}{2}[\Delta_{1}(s_{i}^{z}+s_{i+2}^{z})s_{i+1}%
^{z}+\frac{1}{2}[s_{i+1}^{-}(s_{i}^{+}+s_{i+2}^{+})$ $+h.c.]+[\Delta_{2}%
s_{i}^{z}s_{i+2}^{z}+\frac{1}{2}(s_{i}^{+}s_{i+2}^{-}+s_{i}^{-}s_{i+2}^{+})]$,
and $h_{i,i+1,i+2}^{mag}=\frac{b}{2+x}(s_{i}^{z}+xs_{i+1}^{z}+s_{i+2}^{z}).$
$h_{i,i+1,i+2}^{0}$ is the local Hamiltonian of exchange interaction, and
$h_{i,i+1,i+2}^{mag}$ is the that of applied magnetic field. We have set
$b=\frac{B}{J_{2}},f=\frac{J_{1}}{J_{2}}$ and made scaling $J_{2}=1$. In
$h^{mag}$ we have introduced a free parameter $x$ to facilitate later
calculation. Note that the Hamiltonian is independent of the value of $x$ if
we apply the periodic boundary condition.\newline\textbf{The quantum critical
surface from local site Hamiltonian:} The local Hamiltonian $h_{123}$, which
serves as our starting point, can be diagonalize and the eigenvalues in terms
of parameters $x$, $\Delta_{1,2}$ and $f$ are
\begin{align*}
E_{0}  &  =\frac{1}{4}(-2b+\Delta_{2}+\Delta_{1}f),\\
E_{1}  &  =\frac{1}{4}(2b+\Delta_{2}+\Delta_{1}f),\\
E_{2}  &  =\frac{1}{4}(-2-2b-\Delta_{2})+\frac{b}{2+x},\\
E_{3}  &  =\frac{1}{4}(-2-\Delta_{2}+\frac{2bx}{2+x}),\\
E_{4\pm}  &  =\frac{4b-(-2+\Delta_{1}f)(x+2)\pm\Gamma_{+}}{8(x+2)},\\
E_{5\pm}  &  =\frac{-4b-(-2+\Delta_{1}f)(x+2)\pm\Gamma_{-}}{8(x+2)},
\end{align*}
where
\begin{align*}
\Gamma_{\pm}^{2}  &  =[16b^{2}(-1+x)^{2}\pm8b(2-2\Delta_{2}+\Delta
_{1}f)(-2+x+x^{2})\\
&  +4(-1+\Delta_{2})^{2}-4\Delta_{1}(-1+\Delta_{2})f+(8+\Delta_{1}^{2})f^{2}]
\end{align*}
We found that $E_{0}(E_{1})$ corresponds to the state $\left\vert
\downarrow\downarrow\downarrow\right\rangle (\left\vert \uparrow
\uparrow\uparrow\right\rangle )$, $E_{2}$ and $E_{5\pm}$ correspond to one
spin-up states and $E_{3}$ and $E_{4\pm}$ to the one spin-down states.\newline
In order to see quantum critical points for entire system, we seek in the
spectrum the level crossing between FP state $|\downarrow\downarrow
\downarrow\rangle$ and other eigen state. We need at least two other states
because for an infinite spin chain with space inversion symmetry any other
states must be two-fold degenerate. Inspecting the eigenvalues, we found that
$E_{0},E_{2}$ and $E_{5-}$ are three lowest possible energies. By requiring
$E_{0}=E_{2}=E_{5-}$, we get the following relations
$x=\frac{f(f+4\Delta_{1})}{2(2+2\Delta_{2}+\Delta_{1}f)}$, $b=\frac{1}%
{4}(2+2\Delta_{2}+\Delta_{1}f))(2+x)$
or for indication of the quantum critical surface
\begin{equation}
b=1+\Delta_{2}+\Delta_{1}f+\frac{f^{2}}{8}. \label{QCS}%
\end{equation}
The correspond energy of these three degenerate state is
\begin{equation}
E_{0}=\frac{-2-\Delta_{2}-\Delta_{1}f}{4}-\frac{f^{2}}{16}%
\end{equation}
\textbf{Matrix formulation of the global eigen state:} Here, we generalize the
results of local Hamiltonian for arbitrary number of spins. The global
Hamiltonian of $N$\ spins can be obtained by the direct sum of local
Hamiltonian.
\begin{align*}
H  &  =h\otimes\hat{1}_{2^{N-3}}\oplus\hat{1}_{2^{1}}\otimes h\otimes\hat
{1}_{2^{N-4}}\oplus\hat{1}_{2^{2}}\otimes h\otimes\hat{1}_{2^{N-5}}\\
&  \oplus\dots\oplus\hat{1}_{2^{N-4}}\otimes h\otimes\hat{1}_{2^{1}}\oplus
\hat{1}_{2^{N-3}}\otimes h,
\end{align*}
where each basic element $h$ is just the local Hamiltonian $h_{i,i+1,i+2}$ and
$\hat{1}_{M}$ is the identity matrix of rank $M$. The eigen states of an
infinite spin chain can be also found by adding more spins to eigen states of
the local Hamiltonian. Consider the matrix
\begin{equation}
m_{j}=\left(
\begin{array}
[c]{ccc}%
\left\vert \downarrow\right\rangle _{j} & \left\vert \uparrow\right\rangle
_{j} & 0\\
0 & 0 & \left\vert \downarrow\right\rangle _{j}\\
0 & -\left\vert \downarrow\right\rangle _{j} & -\frac{f}{2}\left\vert
\downarrow\right\rangle _{j}%
\end{array}
\right)  \label{Mn}%
\end{equation}
representing the spin state at site $j$. The three degenerate ground states of
the local Hamiltonian $h_{j,j+1,j+2}$ with energies $E_{0}$, $E_{2}$ and
$E_{5-}$ can be expressed as the matrix $m_{j}m_{j+1}m_{j+2}$. In fact, the
matrix elements are the linear combinations of these three eigen states. It is
clear that $M_{n}=\Pi{_{j}}_{{=1}}^{N}{m_{j}}$ gives the states of $n$ spins.
$h_{j,j+1,j+2}$, operates only on the product $m_{j}m_{j+1}m_{j+2}$ (whose
elements are its eigen states,) without affecting other matrices. Hence, the
operation of Hamiltonian will not alter the form of $M_{n}$ and its matrix
elements are linear combinations of the eigen states. We can write $M_{n}$ in
a more compact form
\[
M_{n}=\left(
\begin{array}
[c]{ccc}%
\left\vert \phi_{n}^{a}\right\rangle  & \left\vert \phi_{n}^{b}\right\rangle
& \left\vert \phi_{n}^{c}\right\rangle \\
0 & -f_{n-1}\left\vert \phi_{n}^{a}\right\rangle  & f_{n}\left\vert \phi
_{n}^{a}\right\rangle \\
0 & -f_{n}\left\vert \phi_{n}^{a}\right\rangle  & f_{n+1}\left\vert \phi
_{n}^{a}\right\rangle
\end{array}
\right)
\]
where $\left\vert \phi_{n}^{a}\right\rangle =|0\rangle$,$\left\vert \phi
_{n}^{b}\right\rangle =\sum_{j=1}^{n}f_{n-j+1}|j\rangle$ and $\left\vert
\phi_{n}^{c}\right\rangle =\sum_{j=1}^{n}f_{n-j}|j\rangle$, with
$|0\rangle=|\downarrow\downarrow\downarrow\downarrow\dots\rangle
,|j\rangle=s_{j}^{+}|0\rangle$ and the coefficients $f_{j}$ satisfying the
relations $f_{0}=0,f_{1}=1,f_{j}+\frac{f}{2}f_{j+1}+f_{j+2}=0$. Note that
$f_{j}=cos(jk)$ where
\begin{equation}
k=cos^{-1}(\frac{-f}{4}), \label{k}%
\end{equation}
satisfies above equations. The first eigen state is the fully polarized (FP)
state and the other two can be expressed as the linear combination of two
independent one-magnon states:
$|\phi_{n}^{b^{\prime}}\rangle=\frac{1}{\sqrt{n}}\sum_{j=1}^{n}e^{ikj}%
s_{j}^{+}|0\rangle$, $|\phi_{n}^{c^{\prime}}\rangle=\frac{1}{\sqrt{n}}%
\sum_{j=1}^{n}e^{-ikj}s_{j}^{+}|0\rangle$
where $j$ is the site position. The periodic boundary condition implies that
$Nk=2m\pi$. Thus we have constructed the eigen states for a zigzag spin chain
of arbitrary length at the surface Eq.(\ref{QCS}).
\newline\textbf{Proof of the quantum critical surface by positive
semi-definite matrix theorem:} We use the theorem of positive semi-definite
matrix to show that for spin chain of arbitrary length, the above mentioned
three degenerate eigen states are actually the global ground state in certain
region. The theorem of positive semi-definite matrix is the following:
\textit{The necessary and sufficient condition for a real symmetric matrix}
$A$ \textit{to be positive semi-definite is} $x^{T}Ax\geq0$ \textit{for all
real vectors} $x$. \textit{If} $M$ \textit{and} $N$ \textit{are positive
semi-definite, then the sum} $M+N$, \textit{and the direct sum} $M\oplus N,$
\textit{and direct product} $M\otimes N$ \textit{are also positive
semi-definite.}

\begin{figure}[t]
\begin{center}
\includegraphics[scale=0.35]{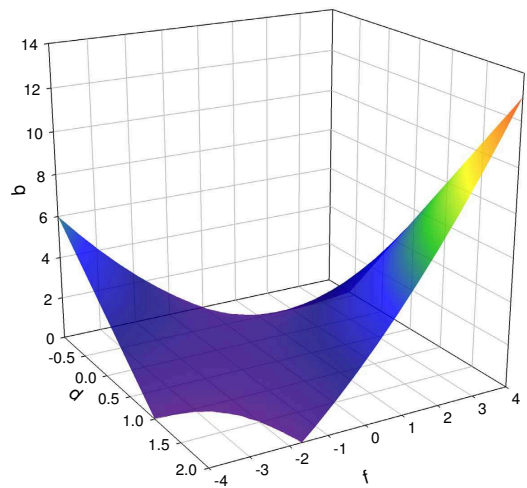}
\includegraphics[scale=0.35]{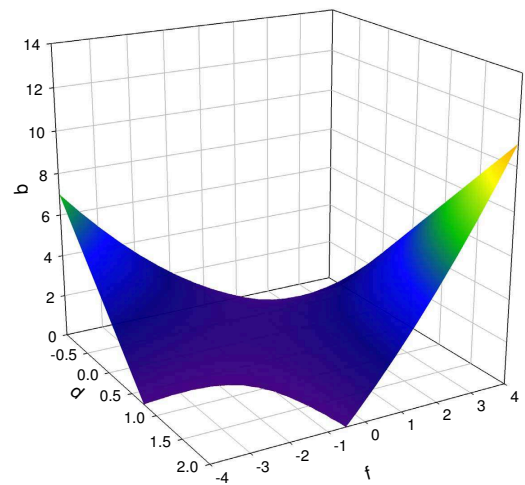}
\end{center}
\caption{The quantum critical surface in the case (a) $\Delta_{1}=\Delta
_{2}=d$, and (b)$\Delta_{1}=d,\Delta_{2}=0$.}%
\label{3Dview}%
\end{figure}

The FP state and the one-magnon states are the degenerate ground states, if
the matrix $(h_{j,j+1,j+2}-E_{0}\hat{1}),$ with $E_{0}$ being their energy, is
a positive semi-definite matrix. We begin the proof with a local Hamiltonian
with three spins. By setting $x=\frac{f(f+4\Delta_{1})}{2(2+2\Delta_{2}%
+\Delta_{1}f)},$ Eq.(\ref{QCS}) and writing the Hamiltonian $h_{i,i+1,i+2}%
-E_{0}$ in a real and symmetric matrix form, we proceed to prove it is a
positive semi-definite matrix. The value $\mathbf{x}^{T}(h_{i,i+1,i+2}%
-E_{0})\mathbf{x}$ where $\mathbf{x}=(x_{1},x_{2},x_{3},x_{4},x_{5}%
,x_{6},x_{7},x_{8})^{T}$ is any real column matrix, is
\begin{align}
&  x^{T}(h_{i,i+1,i+2}-E_{0})x=\frac{1}{4}[4(1+\Delta_{2}+\Delta_{1}%
f+\frac{f^{2}}{8})x_{1}^{2}\nonumber\\
&  +\frac{f(4\Delta_{1}+f)}{2}(x_{2}^{2}+x_{5}^{2})+(4+4\Delta_{2}-\frac
{f^{2}}{2})x_{3}^{2}\nonumber\\
&  +2(x_{2}+\frac{f}{2}x_{3}+x_{5})^{2}+2(x_{4}+\frac{f}{2}x_{6}+x_{7})^{2}].
\label{eq-psd}%
\end{align}
$(h_{j,j+1,j+2}-E_{0}\hat{1})$ is a positive semi-definite matrix provided two
inequalities below are satisfied simultaneously
\begin{equation}
f(4\Delta_{1}+f)\geq0,\quad4+4\Delta_{2}-\frac{f^{2}}{2}\geq0. \label{shade}%
\end{equation}
Since all the local Hamiltonian $(h_{j,j+1,j+2}-E_{0}\hat{1})$ are positive
semi-definite matrices, so is $h_{123}\otimes\hat{1}_{2}+\hat{1}_{2}\otimes
h_{234}$ and $H-E_{0}$. Thus we conclude that in the region given by
inequalities (\ref{shade}), the degenerate ground state is the FP state and
one magnon state with $k$ in Eq.(\ref{k}). The degenerate ground states are
$|\phi_{N}^{a}\rangle=\left\vert 0\right\rangle $ , $|\phi_{N}^{b}\rangle
=\sum_{m=1}^{N}e^{ikm}s_{m}^{+}\left\vert 0\right\rangle $ \newline$|\phi
_{N}^{c}\rangle=\sum_{m=1}^{N}e^{-ikm}s_{m}^{+}\left\vert 0\right\rangle $.
We will give argument below that even if outside of the region given by
inequalities (\ref{shade}), they are still the degenerate ground states. In
the mean time we show the quantum critical surfaces given by Eq.(\ref{QCS})
for the case $\Delta_{1}=\Delta_{2}=d$ in Fig.\ref{3Dview}(a) and the case
$\Delta_{1}=d$ and $\Delta_{2}=0$ in Fig.\ref{3Dview}(b). In the region above
the surfaces only the FP state is the ground state. There are several
interesting features. Firstly, larger $\Delta_{1}$ and $\Delta_{2}$, requires
larger magnetic field $b$. This means that the system is made difficult to be
polarized by anisotropy. On the other hand, negative $f$, ferromagnetic
coupling, reduces the value of $b$ as expected. Eq.(\ref{QCS}) is valid for
$f\geq-4$. Beyond which only the FP state is the ground state. Quantitatively,
we see a linear dependence of $b$ on $\Delta_{1}$ and $\Delta_{2}$. This
indicates that treating the coupling $S_{i}^{z}S_{j}^{z}$ with the mean-field
approximation is accurate at least in the region close to that given by
Eq.(\ref{QCS}).\newline
\newline\textbf{Analytical solutions outside the region of Eq.(\ref{shade})):}
Beyond the regions given by (\ref{shade}), the global Hamiltonian is still
likely to be positive semi-definite. We apply Bethe ansatz to analyze this
region. Our approach is the following. We write down the Bethe ansatz solution
for $n$ up-spins state, and the constraints of the coefficients. For a given
$b$ and a given$\ f$, we find the state with lowest energy. We are able to
show that for $n\leq\frac{N}{3}$, the energy of above mentioned state is
greater or equal to the energy of FP state on the surface given by
Eq.(\ref{QCS}). We begin by writing the state as the linear combinations of
$n$ spin-up states:
\begin{equation}
\left\vert \psi\right\rangle =\sum_{x_{1}<x_{2}<\dots<x_{n}}a_{x_{1}%
,x_{2},\dots,x_{n}}s_{x_{1}}^{+}s_{x_{2}}^{+}\dots s_{x_{n}}^{+}\left\vert
0\right\rangle \label{psi}%
\end{equation}
where $a_{x_{1},x_{2},\dots,x_{n}}$ is symmetric under exchanges of spin
indices. Their constraints can be obtained by requiring $\left\vert
\psi\right\rangle $ to be the eigen state of the Hamiltonian.\newline In this
case there are many constraints. But one of them has the property
$x_{i+1}>x_{i}+2$ with all possible $i$. This means that all the up-spins are
separated by at least two down-spins, or no two up-spins are NN or NNN. This
relation is
\begin{align*}
&  2(E-E_{0})a_{x_{1},x_{2},\dots,x_{n}}\\
&  =2n(b-\Delta_{2}-\Delta_{1}f)a_{x_{1},x_{2},\dots,x_{n}}\\
&  +f\sum_{i=1}^{n}(a_{\dots x_{i-1},x_{i}-1,x_{i+1\dots}}+a_{\dots
x_{i-1},,x_{i}+1,x_{i+1\dots}})\\
&  +\sum_{i=1}^{n}(a_{\dots x_{i-1},x_{i}-2,x_{i+1\dots}}+a_{\dots
x_{i-1},,x_{i}+2,x_{i+1\dots}})
\end{align*}
With the same method for two-magnon states, we let $a_{x_{1},x_{2},\dots
,x_{n}}=\sum_{P}A_{P}\exp[i\sum_{j=1}^{n}k_{Pj}x_{j}]$, where $P$ means
permutation of spin sites $j$. The equation becomes
\begin{equation}
2(E-E_{0})=2n(b-\Delta_{2}-\Delta_{1}f)+2f\sum_{i=1}^{n}\cos k_{i}+2\sum
_{i=1}^{n}\cos2k_{i}. \label{energy}%
\end{equation}
If $k_{i}$'s were independent variables, then the lowest energy would have
been obtained by varying them independently. The result is
\begin{equation}
k_{i}=k=\pm\cos^{-1}(-\frac{f}{4}). \label{ki}%
\end{equation}
For $E=E_{0}$ and Eq.(\ref{ki}), we again reach the relation in Eq.
(\ref{QCS}). Since $k_{i}$'s have other constraints, the energy we thus get is
the lower bound. Hence, we conclude that for $1\leq n\leq\frac{N}{3}$, the FP
state and one-magnon states have the lowest energy with the magnetic field
given by Eq. (\ref{QCS}). It is plausible to assume that this is true for
arbitrary value of $n$.
Bethe ansatz gives the excitation energy of magnons as $\Delta E=E-E_{0}%
=1+\frac{f^{2}}{8}+f\cos k+\cos2k$. Interestingly, it does not depend on the
anisotropy $\Delta_{1}$ and $\Delta_{2}$. It vanishes when Eq.(\ref{k}) is
satisfied. Physically, this means that the magnons soften at certain wave
vectors, depending on the value of $f$. This phenomena can be detected by
neutron scattering or susceptibility measurement.\newline\textbf{Application
to superconducting quantum dot system:} Here we derive the analytical
expression of the quantum critical surface of a superconducting quantum dots
(SQD) array with NN and NNN tunneling and Coulomb interaction in the presence
of gate voltage in every superconducting dot. The quantum critical surface
separate the charge density wave phase of lower commensurability ($n=1/2$),
which corresponds to a anti-ferromagnetic phase of a spin chain; from a higher
commensurability state ($n=1$, or integer numbers of Cooper pair in every
dot,) which correspondence to the FP phase. Therefore, we use the relation
between $b$ and $f$ evaluated in the previous section to study the quantum
critical surface between these two phases. The quantum fluctuations are
controlled by the system parameters like charging energy of SQD and the
Josephson coupling ($E_{J}$). In the SQD model \cite{sar1}, we consider the
finite range $E_{J}$ and Coulomb interaction which have experimental support
\cite{havi1,zant,havi2,havi3}. The model Hamiltonian of our study consists of
different interactions,
$H=H_{J1}+H_{J2}+H_{EC0}+H_{EC1}+H_{EC2}$.
We recast our basic Hamiltonians in the spin language around the charge
degeneracy point.
$H_{J1}=2E_{J1}\sum_{i}({S_{i}}^{\dagger}{S_{i+1}}^{-}+h.c)$, $H_{J2}%
=2E_{J2}\sum_{i}({S_{i}}^{\dagger}{S_{i+2}}^{-}+h.c)$,
$H_{EC0}=\frac{E_{C0}}{2}\sum_{i}{(2{S_{i}}^{Z}+h)^{2}}\approx{2hE_{C0}}%
\sum_{i}{2{S_{i}}^{Z}}$ , $H_{EC1}=4E_{Z1}\sum_{i}{S_{i}}^{Z}~{S_{i+1}}^{Z}$,
$H_{EC2}=4E_{Z2}\sum_{i}{S_{i}}^{Z}~{S_{i+2}}^{Z}$.
where $H_{J1}$ and $H_{J2}$ are Josephson energy Hamiltonians respectively for
NN and NNN Josephson tunneling between the SQD and ${H_{EC0}},{H_{EC1}}$, and
$H_{EC2}$ are respectively the Hamiltonians for on-site, NN and NNN charging
energies of the SQD. $h=\frac{2n+1-N}{2}$ is a parameter which allows tuning
the system to a degeneracy point by means of gate voltage. When ${E_{C0}%
}>>{E_{J}}$, sequential tunnelling of Cooper pair across the SQD is not a
energetically favorable process. One must consider the co-tunneling effect,
i.e., the higher order expansion in $\frac{E_{J1}}{E_{C0}}$, which reduce the
intersite Coulomb charging energy and also increase the NNN $E_{J}$
\cite{sar1,lar,giu}. This results in new tunneling energies: ${E_{J2}%
\rightarrow E_{J2}}+{{E}_{J1}}^{2}/2E_{C0},$ and ${E_{Z1}}\rightarrow
E_{Z1}-3{E_{J1}}^{2}/16E_{C0}$.\newline Now we present the analytical
expression of quantum critical surface for the presence and absence of
co-tunneling effect, following the relation $b$ and $f$ in Eq.(\ref{QCS}). The
analytical expression of quantum critical surface in the absence of
co-tunneling effect are the following\newline$2h{E_{C0}}={2{E_{J2}}%
}+{4({E_{z1}}+{E_{z2}})}+\frac{{{E}_{J1}}^{2}}{4{E_{J2}}}$\newline while that
in the presence of co-tunneling effect is\newline$2hE_{C0}=2E_{J2}%
+4(E_{z2}+E_{z1})+\frac{E_{J1}^{2}}{4}+\frac{E_{J1}^{2}}{4E_{J2}}%
+\frac{2E_{J1}^{2}}{E_{C0}}.$
We have already calculated the quantum critical surface on which FP state and
one magnon state are degenerate. Here the one vacant site of Cooper pair in
the SQD lattice array corresponds to one magnon state and Mott insulating
state with integer numbers of Cooper pairs in each lattice site corresponds to
the FP state. The applied gate voltage on each SQD of the array corresponds to
the applied magnetic field. The charging energy corresponds to the anisotropy
of the exchange interaction of the spin system. For large gate voltage, the
SQD is in the insulating state. Interestingly, the charging energy tends to
reverse this trend by destabilizing the state of integer numbers of Cooper
pairs in each lattice site and makes the system conducting. The analytical
expression for the quantum critical surface for the SQD array is the first in
the literature.\newline\textbf{Conclusions:} We have found the quantum
critical surface of a zigzag spin chain under magnetic field by using the
theorem of positive semi-definite matrix and Bethe-ansatz methods. We have
also find the quantum critical surface of superconducting quantum dots system
by using our analytical methods.

\end{document}